\documentclass[12pt]{iopart}

\usepackage{iopams}
\usepackage{amssymb}
\usepackage{graphicx}
\eqnobysec

\begin{document}

\title[Local phase structure of wave dislocation lines]{Local phase structure of wave dislocation lines: twist and twirl}

\author{M R Dennis}

\address{H H Wills Physics Laboratory, Tyndall Avenue, Bristol BS8 1TL, UK}

\begin{abstract}
Generic wave dislocations (phase singularities, optical vortices) in three dimensions have anisotropic local structure, which is analysed, with emphasis on the twist of surfaces of equal phase along the singular line, and the rotation of the local anisotropy ellipse (twirl). Various measures of twist and twirl are compared in specific examples, and a theorem is found relating the (quantised) topological twist and twirl for a closed dislocation loop with the anisotropy C line index threading the loop.
\end{abstract}


\section{Introduction}\label{sec:int}

Phase singularity lines in 3-dimensional scalar waves, that is, nodal lines or wave dislocations \cite{nb:34, nye:natural}, are often called (optical) vortices since they are vortices of current, or energy, flow. 
Being places where the phase is singular, all phases are found in the vortex neighbourhood, and the phase change is arbitrarily fast there. 
The surfaces on which phase is constant all meet on the singularity line, and are often twisted around the singularity as helicoids, as with the familiar screw dislocations \cite{nb:34}; it is this twisting that distinguishes screw dislocations from edge dislocations, whose phase structure does not rotate along the dislocation line.

In two dimensions, phase singularities are generically points where the intensity of the wave vanishes and the phase is singular. 
Knowledge of the phase contour lines near the point singularities can be sufficient to fill in the phase structure of the rest of the field. 
In 3-dimensional fields, topological singularities form a network of lines in space, often described as a `skeleton' of the spatial pattern  \cite{nye:natural, hll:topology} - in complex scalar fields, the phase structure local to the phase singularity lines show how the global phase field is constructed upon this skeleton. 
The simplest 3-dimensional property of a phase singularity is its sense, or {\itshape topological current}: the direction endowed on the singularity line by the right-handed increase of phase. 
The topological current direction is preserved along the singularity line. 
The sense of the phase helicoids twisting around the singularity is independent of this; for instance (as demonstrated below), if a dislocation is embedded in a plane wave whose propagation direction is parallel to the topological current direction, the phase surfaces ending on the singularity form left-handed helicoids; if the propagation is antiparallel to the dislocation sense, they are right-handed. 
The twist is rather more complicated in more general situations, such as: when the singularities evolve off-axis in gaussian beams, with a noncanonical transverse shape \cite{ss:parameterization, mwt:noncanonical}; when they are knotted, linked, or braided \cite{bd:332, bd:333, dennis:braided}; in isotropic random plane waves, when the singularities are tangled in a nontrivial way \cite{bd:321}; when they form the characteristic `antelope horns' of the elliptic umbilic diffraction catastrophe \cite{bnw:79}.
The zero contour surfaces of the real or imaginary parts of the field are often used to locate phase singularities and describe their geometry \cite {bd:321, freund:trajectories}; these are, of course, special cases of the general phase contours.

\begin{figure}
\begin{center}
\includegraphics*[width=10cm]{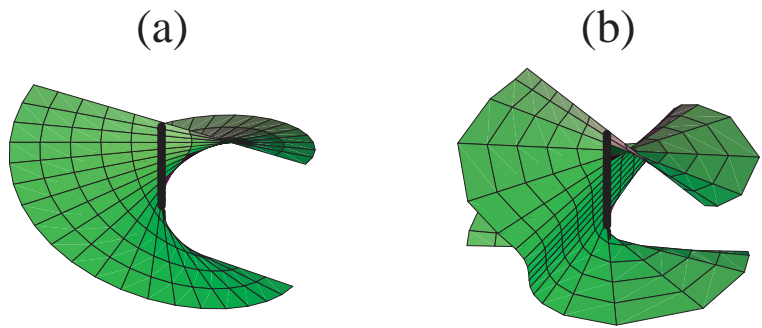}
\end{center}
    \caption{Two surfaces of constant phase (mod $\pi$) in the vicinity of a twisted wave dislocation. (a) Uniform twist. This is a surface of constant phase (mod $\pi$) of the simple screw dislocation (\ref{eq:edgescrew}) with $\bi{k} = (0,0,k),$ and is a uniform helicoid. (b) Nonuniform twist. This constant phase surface is an example of the generic situation, calculated from (\ref{eq:twex}) with $\beta = 2\alpha$ and $c = -1/2.$}
    \label{fig:surf}
\end{figure}

My aim here is to describe this 3-dimensional phase twist structure in the general case, illustrated by simple examples of dislocated waves. 
As these examples show, it is very easy to find dislocations in waves that do not have a uniform phase screw structure, and this situation is, in fact, what one would expect generically.
The main complication to the description comes from the fact that different phase helicoids twist at different rates, and this must somehow be averaged to give an overall sense of dislocation twist at a point on the singularity line. 
This is related to the fact that the local phase structure transverse to a dislocation line is generically anisotropic, squeezed into an ellipse \cite{ss:parameterization, mwt:noncanonical, bd:321, dennis:thesis, berry:330, dennis:local}, complicating the averaging around the dislocation. 
An example of two phase helicoids near a dislocation are shown in figure \ref{fig:surf}; the first has a uniform helicoidal structure, the other does not.

The notation of \cite{bd:321, dennis:thesis} will be followed. 
In particular, $\psi = \psi(\bi{r})$ denotes a complex scalar wave, and in terms of real and imaginary parts, or amplitude and phase,
\begin{equation}
   \psi = \xi + \rmi \eta = \rho \exp(\rmi \chi).
   \label{eq:psi}
\end{equation}
Time dependence will not be considered here, and terms such as twist and twirl refer to variation with distance, not time.
Local cartesian coordinates $\bi{r} = (x,y,z)$ or cylindrical coordinates $R,\phi,z$ will be used, with the dislocation along $x=y=r=0,$ and phase $\chi$ increasing with $\phi$ (that is, the topological current is parallel to the $+z$-direction). 
Important formulae will be given in coordinate-free form, with the dislocation tangent denoted by $\bi{T}$ (which is $(0,0,1)$ in local coordinates), and the directional derivative along the dislocation by $\bullet '$ (which is $\partial_z \bullet$ in local coordinates). 
Several of the results described here were also discussed in \cite{dennis:thesis}, particularly in section 2.8, pages 54-58.

The epitome of a twisted wave dislocation is in the example \cite{nb:34}
\begin{equation}
   \psi_0 = (x + \rmi y) \exp(\rmi \bi{k}\cdot\bi{r}).
   \label{eq:edgescrew}
\end{equation}
This wave is a local solution of the Helmholtz equation $\nabla^2 \psi + k^2 \psi = 0.$
It is a {\itshape screw dislocation} when $\bi{k} = (0,0,k),$ and each of the surfaces of constant phase is a regular helicoid with pitch $2\pi/k,$ reminiscent of the atomic planes in a crystal lattice near a crystal screw dislocation \cite{read:dislocations}. 
It may be compared with an {\itshape edge dislocation}, where in (\ref{eq:edgescrew}), $\bi{k} = (0,k,0).$ 
In this case, the phase structure does not change with $z,$ and the phase lines have a structure similar to the arrangement of the atomic planes near a crystal edge dislocation \cite{read:dislocations}. 
A general $\bi{k}$ in (\ref{eq:edgescrew}) gives a {\itshape mixed edge-screw dislocation.} 
Far from the dislocation, the phase of (\ref{eq:edgescrew}) has a plane wave character with wavevector $\bi{k}.$

This example, and its crystal analogy, motivated Nye \cite{nye:motion} to define the {\itshape Burgers vector} of a wave dislocation
\begin{equation}
   \bi{b} = \left(\lim_{x\to 0} \left(\chi_x\right)_{y=z=0}, \lim_{y\to 0}
   \left(\chi_y\right)_{x=z=0}, \lim_{z\to 0} \left(\chi_z\right)_{x=y=0}\right),
   \label{eq:burgers}
\end{equation}
where subscripts after brackets denote quantities held constant, other subscripts denote partial derivatives, and coordinates are local. 
The dislocation (\ref{eq:edgescrew}) has Burgers vector $\bi{b} = \bi{k},$ which is the wavevector of the asymptotic plane wave. 
This gives the expected results of the vector being parallel and perpendicular to the screw and edge dislocation lines respectively. 
However, there is a danger of taking the analogy between wave dislocations and crystal dislocations too far, and there are problems with this definition of a Burgers vector for a wave dislocation in other rather general situations.

In the derivation of (\ref{eq:burgers}) in \cite{nye:motion}, it is assumed that the dislocation is curved and moving (i.e. the wave is narrow-band); for monochromatic waves, the dislocation is stationary, and many dislocation lines occurring in optics are straight lines (often parallel to the beam direction). 
If the only available information about the field is local to the dislocation, for instance, a finite Taylor expansion of (\ref{eq:edgescrew}) about $x = y = 0,$ $b_z$ is not defined.
In more general cases of wave interference, such as isotropic random plane wave superpositions \cite{bd:321}, there is no overall propagation direction, and only local properties of the dislocation, determined using derivatives of $\psi$ on the dislocation, are relevant.

It is desirable to have a measure of screwness, determining the twist of the local phase structure along the dislocation, in terms only of local derivatives.
The following is an exploration of this twist geometry, and its realisation in simple beam superpositions.

\section{The geometry of twist and twirl}\label{sec:geometry}

The {\itshape twist} $Tw$ is defined to be the rate of rotation of phase along a dislocation, according to several different measures to be described. 
The simplest geometric twist is the rotation of a surface of constant phase along the dislocation, which is also the rate of rotation of the normal to the phase surface. 
For convenience, a surface of constant phase in the vicinity of the phase singularity will be called a {\itshape phase ribbon} \cite{ws:filaments1,ws:filaments2}; for most of the discussion, only the ribbon geometry is important. Each plot (a) and (b) of figure \ref{fig:surf} shows two phase ribbons, with phases differing by $\pi.$
 
Fixing a particular phase $\chi_0,$ the normal to the $\chi_0$-ribbon is
\begin{eqnarray}
   \bi{U}_{\chi_0} & = \rm{Re} \{\nabla \psi \exp(-\rmi \chi_0)\} \nonumber \\
   & = \nabla \xi \cos \chi_0 + \nabla \eta \sin\chi_0.
   \label{eq:udef}
\end{eqnarray}
Using $\phi$ as a local azimuthal coordinate, the rate of twist $Tw(\chi_0)$ of the $\chi_0$-ribbon is
\begin{eqnarray}
   Tw(\chi_0) & = \left(\phi_z\right)_{\chi=\chi_0} \nonumber \\
   & = \partial_z \arctan \frac{U_{\chi_0,y}}{U_{\chi_0,x}} \nonumber \\
   & = \frac{\bi{T}\cdot\bi{U}_{\chi_0} \times \bi{U}_{\chi_0}'}{U_{\chi_0}^2}.
   \label{eq:twch}
\end{eqnarray}
Applying this formula to the wave (\ref{eq:edgescrew}) gives $-k_z.$ Apart from the sign, this is the $z$-component of the Burgers vector.
The $-$ sign originates in the fact that the helicoid is left-handed, a general property of dislocations whose topological current is parallel to the propagation direction \cite{dennis:thesis, dennis:braided}.

\begin{figure}
\begin{center}
\includegraphics*[width=8cm]{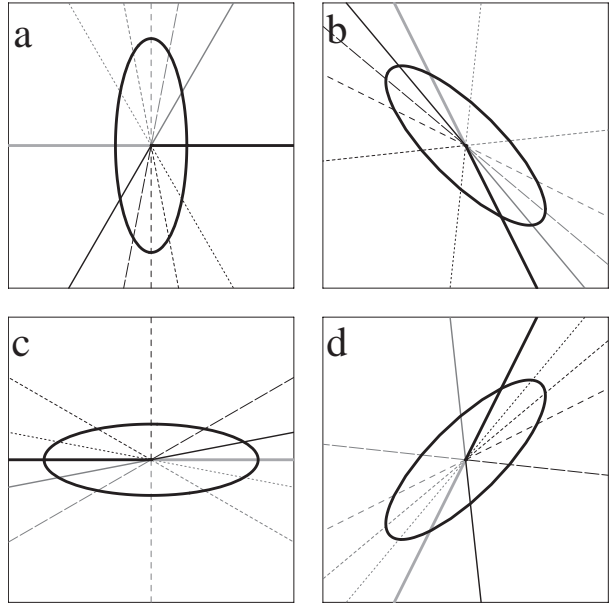}
\end{center}
    \caption{The sequence of transverse phase lines (separated by an equal phase difference $\pi/6$) and anisotropy ellipse of a twisting, twirling dislocation of the form (\ref{eq:twex}), with $\beta = 2\alpha, c = -1/2.$ the $z$-spacing between each frame is $\pi/2\alpha.$ The twist and twirl have opposite senses. The surface in figure (b) is swept out by a pair of opposite phase lines in this figure; the parameters in the two figures are the same.}
    \label{fig:twirltab}
\end{figure}

The phase structure transverse to the dislocation in (\ref{eq:edgescrew}) is isotropic, and all the phase ribbons twist at the same rate as each other, and along the dislocation (one such is shown in figure \ref{fig:surf} (a)). 
General dislocations are anisotropic, however, and the local contours of intensity $\rho^2$ are elliptical, with a corresponding squeezing of phase gradient $\nabla\chi.$ 
It can be shown \cite{bd:321, dennis:thesis, dennis:local} that $(\chi_{\phi})_{z=0} = R^2 \omega /\rho^2,$ where $\omega$ is the vorticity on the dislocation, defined as $\omega \equiv |\nabla \xi \times \nabla \eta | = |\nabla \psi^{\ast} \times \nabla \psi|/2$ (the direction of this vector gives the topological current). 
The plots in figure \ref{fig:twirltab} indicate this aspect of elliptic phase squeezing.
In fact, both the ellipse and the phase squeezing are accounted for by the gradient vector $\nabla \psi$ on the dislocation. 
As with all complex vectors, it is associated with an ellipse, traced out by $\bi{U}_{\chi_0}$ as $\chi_0$ changes; this ellipse is the same shape as that described by $\rho^2$ and $\nabla \chi,$ but with axes exchanged.

It is more natural to quantify the twist of the entire singularity core, rather than simply for a fixed phase. 
Therefore, it is necessary to average $Tw(\chi_0)$ over all phases, although it is not clear whether averaging with respect to the phase $\chi$ or azimuth $\phi$ is appropriate. 
It is possible to calculate both averages: the phase average (integrating $Tw(\chi_0)$ with respect to $\chi_0$) has a complicated form involving the axes of the phase ellipse, and is in \cite{dennis:thesis} equation (2.8.6); the azimuth average (integrating $Tw(\chi_0)$ with respect to $\phi$) has a simpler form, and is (\cite{dennis:thesis}, equation (2.8.7))
\begin{equation}
   Tw_{\phi} = \frac{1}{2\pi} \int_0^{2\pi} \rmd \phi \,
   Tw(\chi)  = \frac{\rm{Re}\{\bi{T}\cdot \nabla \psi^{\ast} \times \nabla
   \psi'\}}{2\omega}.
   \label{eq:twph}
\end{equation}
This depends only on derivatives up to the second of $\psi$ on the dislocation line. 
As expected, it gives $-k_z$ for (\ref{eq:edgescrew}).

Berry \cite{berry:330} chose to average phase instead by examining the rate of change of phase at a fixed azimuth $(\chi_z)_{\phi = \phi_0}.$ 
The result of a particular averaging, he defined the {\itshape screwness} $\sigma$ to be
\begin{equation}
   \sigma = \frac{-\rm{Im} \{\nabla \psi^{\ast} \cdot \nabla
   \psi'\}}{|\nabla \psi|^2}.
   \label{eq:screwness}
\end{equation}
($\sigma$ defined here is the negative of that defined in \cite{berry:330}.)
The screwness for (\ref{eq:edgescrew}) is $-k_z.$

The difficulty in defining the total twist arises because the different phase ribbons twist at different rates, due to the phase anisotropy ellipse associated with $\nabla \psi.$ 
Along the dislocation line, the anisotropy ellipse itself may rotate, as well as change its size and eccentricity. 
Because the rotation of the ellipse is independent of the phase twist, it will be referred to as the {\itshape twirl} $tw$ of the dislocation line, and may be found as follows.

The complex vector field $\nabla \psi$ shares geometric features associated with vector polarization fields \cite{dennis:thesis,bd:324} \footnote{The two fields do not have identical structures, since free field vector solutions of Maxwell's equations are divergence free, whereas $\nabla \psi,$ being a gradient field, is curl free.}. 
In particular, in local coordinates, $\nabla \psi$ is confined to the $xy$-plane, and therefore parameters, describing all the geometric properties of the ellipse, may be defined:
\begin{eqnarray}
   S_0 = |\nabla \psi|^2, \qquad & S_1 = |\psi_x|^2 -|\psi_y|^2, \nonumber \\
   S_2  = \psi_x^{\ast} \psi_y + \psi_x \psi_y^{\ast}, \qquad  &
   S_3  = -\rmi (\psi_x^{\ast} \psi_y + \psi_x \psi_y^{\ast}) =
   2\omega.
   \label{eq:stokes}
\end{eqnarray}
These parameters describing the anisotropy are analogous to the Stokes parameters in polarization; the anisotropy ellipse is related to $\nabla \psi$ and these parameters as the polarization ellipse is related to the electric vector and the Stokes parameters. 
The parameters (\ref{eq:stokes}) do not, themselves, have anything to do with polarization.
The anisotropy ellipse, depending only on $\nabla\psi,$ may be defined at any point of the scalar field, not only on a dislocation; the ellipse is circular (or linear) generically along lines in space \cite{bd:324, nh:wavestructure, dennis:thesis}.

The azimuthal angle of orientation of the major ellipse axis is $\arg(S_1 + \rmi S_2)/2.$ 
The twirl $tw_{\phi}$ may therefore be defined as the rate of change of this angle along the dislocation line:
\begin{equation}
   tw_{\phi} = \frac{1}{2} \frac{S_1 S_2' - S_2 S_1'}{S_1^2 +
   S_2^2}.
   \label{eq:twphi}
\end{equation}
The denominator is equal to $S_0^2 - S_3^2 = |\nabla \psi|^4 - 4 \omega^2,$ which is zero when the ellipse is circular and twirl is not defined.

The natural measure of phase around the ellipse associated with a complex vector is the {\itshape rectifying phase} $\chi_{\rm{r}}$ \cite{dennis:thesis, dennis:polarization} (for polarization ellipses, it also called phase of the vibration \cite{nye:natural}). 
$\chi_{\rm{r}}$ is defined such that the complex vector $\exp(-\rmi \chi_{\mathrm{r}}) \nabla \psi$ has orthogonal real and imaginary parts (the real part along the ellipse major axis, the imaginary along the minor), and can be shown \cite{dennis:polarization} to be equal to $\arg (\nabla \psi \cdot \nabla \psi)/2.$ 
The {\itshape phase twirl} $tw_{\chi}$ may be defined as the rate of change of this phase along the dislocation:
\begin{equation}
   tw_{\chi} = \frac{1}{2} \left(\arctan \frac{2 \nabla \xi \cdot
   \nabla \eta}{|\nabla \xi|^2 - |\nabla \eta|^2}\right)'.
   \label{eq:twchi}
\end{equation}
This gives a natural measure of the rate of change of phase with respect to the ellipse axes. Its form is not particularly simple when the derivative in (\ref{eq:twchi}) is taken, although it is easily seen that the denominator is $|\nabla \psi|^4 - 4 \omega^2.$ This implies that $tw_{\chi}$ is not defined when the ellipse is circular (isotropic).

The two twirls here defined in (\ref{eq:twphi}), (\ref{eq:twchi}) may be combined to give a new measure of the phase twist. 
Since $tw_{\chi}$ measures the rate of change of phase with respect to the ellipse axes, its negative gives a sense for the helicoid phase twist with respect to the ellipse. 
Therefore, the difference $tw_{\phi} - tw_{\chi}$ gives an {\itshape ellipse-defined twist} $Tw_{\rm{ell}},$ which can be shown to be
\begin{equation}
   Tw_{\rm{ell}}  = tw_{\phi} - tw_{\chi} = \frac{\rm{Re}\{ \bi{T} \cdot(\nabla \psi^{\ast} \wedge \nabla
   \psi') + \rmi \nabla \psi^{\ast}\cdot \nabla \psi'\}}{|\nabla \psi|^2
   +2\omega}.
   \label{eq:twell}
\end{equation}
Although neither type of twirl is defined when the ellipse is circular, $Tw_{\rm{ell}}$ is, and for the example (\ref{eq:edgescrew}), it is $-k_z,$ as desired. 
It is also interesting to note that $Tw_{\rm{ell}}$ is the sum of the numerators of $Tw_{\phi}$ and $\sigma,$ divided by the sum of the denominators.

A simple example of a twisting, twirling dislocation is found in the sum of two screw dislocated waves, with dislocations in opposite directions, and with different pitches $2\pi/\alpha, 2\pi/\beta:$
\begin{equation}
   \psi_1 = (x + \rmi y) \exp(\rmi \alpha z) + c (x - \rmi y)
   \exp(\rmi \beta z)
   \label{eq:twex}
\end{equation}
with $c$ in general complex, with $1 > |c|^2$ ensuring that the dislocation along $x = y = 0$ has topological current in the $+z$-direction.

(\ref{eq:twex}) could represent, for example, the sum of two copropagating but counterrotating order one Laguerre-Gauss or Bessel beams with different $k_z$ components, in the vicinity of the $z$-axis. 
A possible realisation, in terms of Bessel beams, is $\exp(\rmi \phi+ \rmi \alpha z) J_1(\sqrt{1-\alpha^2} R) + c \exp(-\rmi \phi+ \rmi \beta z) J_1(\sqrt{1-\beta^2} R) \sqrt{(1-\alpha^2)/(1-\beta^2)}$.
The rotation of a tranverse interference pattern along a beam, achieved though superposing beams different axial phase dependencies, was also studied by Courtial \cite{courtial:selfimaging}, where the Gouy phases of two superposed singular beams were different; here, the $k_z$ components of the two dislocated fields are different.

The phase change around the dislocation in (\ref{eq:twex}) is anisotropic for $c \neq 0;$ the Stokes parameters $S_0, S_3$ are constant, indicating the anisotropy ellipse has a constant area $\pi(1-|c|^2)$ and eccentricity $2|c|^{1/2}/(1+|c|).$ 
(\ref{eq:twex}) is therefore a local normal form for dislocations whose twist and twirl are much
greater than their curvature or rate of change of anisotropy.

The two types of twirl are computed to be
\begin{equation}
   tw_{\phi} = -(\alpha - \beta)/2, \quad tw_{\chi} = (\alpha
   +\beta)/2 \; \rm{for} \, \psi_1.
   \label{ex1twirl}
\end{equation}
The various rates of twist are
\begin{equation}
   \left. \begin{array}{ll} Tw_{\rm{ell}} & = -\alpha \\
   Tw_{\phi} & = -(\alpha - |c|^2 \beta)/(1-|c|^2) \\
   \sigma & = -(\alpha + |c|^2 \beta)/(1+|c|^2) \end{array}
    \right\} \, \rm{for} \, \, \psi_1.
   \label{eq:ex1twist}
\end{equation}

For this example of a dislocation with a uniform twirl, and constant anisotropy, the different measures of twist are, in general, different. 
They are equal if $c\to 0,$ (yielding the screw dislocation in (\ref{eq:edgescrew})), or if the twirl is zero, i.e. $\alpha = \beta$ (this was the case discussed in \cite{dennis:thesis}). 
Of the different measures of twist, it appears that $Tw_{\rm{ell}}$ agrees numerically the most with intuition: if $\beta$ is a positive integer multiple $m \alpha$ of $\alpha,$ the ellipse will undergo $m-1$ rotations as the phase undergoes one, and the entire pattern is periodic (as in figure \ref{fig:twirltab}), with period $2\pi/\alpha.$
$Tw_{\rm{ell}}$ is the only measure of twist to reflect this.
Equivalently, it can be argued that since the first term on the right hand side of (\ref{eq:twex}) defines the dislocation direction, it also defines the twist, and the effect of the second term is merely to modulate the pattern to produce the twirl.

\section{Closed dislocation topology}\label{sec:closed}

When phase singularity lines form closed loops, certain topological identities must be satisfied. In relation to twist, continuity of the wavefunction requires that the total number of twists of each phase ribbon must be a (positive or negative) integer, which is the same for each ribbon. 
This {\itshape screw number} is therefore a property of the dislocation loop, and is positive if the topological twisting is right handed with respect to the dislocation direction, negative if left handed. For obvious geometric reasons, a dislocation loop with nonzero screw number will be called a {\itshape closed screw dislocation}.

The importance of the screw number is that it gives the dislocation strength threading the loop, by the {\itshape twisted loop theorem}: the screw number $m$ of a strength 1 dislocation loop is equal to minus the dislocation strength threading the loop (in a right handed sense). 
This result is discussed and proved in \cite{dennis:thesis, ws:filaments2, ws:filaments3, ws:filaments4, bd:332}. 
If the dislocation loop is planar, then the integral of the ribbon twist around the loop divided by $2\pi$ gives the screw number. 
If the loop is nonplanar, then the C\v{a}lug\v{a}reanu-White-Fuller theorem \cite{ws:filaments3} implies that the writhe of the curve must be added to the twist integral.
Only planar curves will be considered here.

A simple wave containing a closed screw dislocation can be made from a combination of polynomial waves in cylindrical coordinates
\cite{bd:333}:
\begin{equation}
   \psi_{\rm{closed}} = R^{|m|} \exp(\rmi m \phi)\exp(\rmi k z) (R^2 -
   R_0^2 + 2 \rmi (|m|+1) z/k).
   \label{eq:clscrew}
\end{equation}
This wave has a closed screw dislocation in the $z=0$ plane at $R=R_0,$ with screw number $m,$ its topological current directed opposite to the increase of $\phi.$ 
It is threaded by a strength $m$ dislocation up the $z$-axis (its sense in $+z$). 
In \cite{bd:332, bd:333}, high-strength loops with similar geometry were found. 
If $m=0,$ the loop is the familiar closed edge dislocation loop. 
$Tw_{\phi}, Tw_{\rm{ell}}$ and $\sigma$ are $-m/R_0$ on the closed loop; the screw number is $m.$ 
The twirl $tw_{\phi}$ is zero for this loop, and the anisotropy ellipse axes are oriented in the $R,z$ directions.

\begin{figure}
\begin{center}
\includegraphics*[width=6cm]{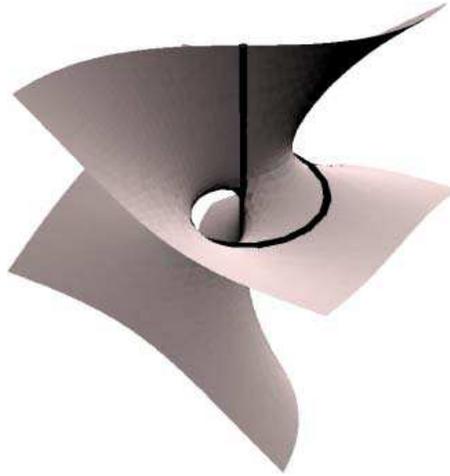}
\end{center}
    \caption{A phase surface (mod $\pi$) of the wave $\psi_{\rm{closed}}$ with $m = 1.$ A straight vertical dislocation line threads a closed screw dislocation, and the local  phase ribbons join up in the form of a noncompact torus.}
    \label{fig:noncompact}
\end{figure}

The nature of the twisted phase ribbons near the dislocation have consequences for the global topology of the total phase surface, that is, the wavefront. 
Figure \ref{fig:noncompact} shows a surface of constant phase (modulo $\pi$) of the wave $\psi_{\rm{closed}}$
with $m = 1.$ 
This surface is a `noncompact torus' (just as the plane is a `noncompact sphere'), extended to infinity because of the infinite straight dislocation on the axis. 
Unlike a compact torus, there is no way of distinguishing the two sides of the surface.
The discussions in \cite{dennis:thesis, ws:filaments2} show that such complex wavefronts are inevitable with closed screw dislocations.

The closed dislocation of (\ref{eq:clscrew}) has zero twirl. 
Waves with closed twirling dislocations loops can be made by superposing waves of the form $\psi_{\rm{closed}}$ with dislocations of opposite sense and different topological twists:
\begin{eqnarray}
   \psi_2 & = R^{|m_1|} \exp(\rmi m_1 \phi)\exp(\rmi k z) (R^2 -
   R_0^2 + 2 \rmi (|m_1|+1) z/k) \nonumber \\
   & \quad + c R^{|m_2|} \exp(\rmi m_2 \phi)\exp(-\rmi k z) (R^2 -
   R_0^2 - 2 \rmi (|m_2|+1) z/k).
   \label{eq:cltwirl}
\end{eqnarray}
When $m_1, m_2$ are different, we may expect the twirl to be nonzero, as with (\ref{eq:twex}). 
Not every combination of $m_1, m_2$ yields the expected twirling wave however, because the complicated threading interference structure is affected in a nonlinear way. 
For simplicity, $k$ and $R_0$ are taken as 1.

Choosing $m_1 = 1, m_2 = 0, c = 2/3$ gives a wave with a twirling dislocation, and the measures of twist are
\begin{equation}
\fl   \left. \begin{array}{ll} Tw_{\rm{ell}} & = -1 \\
   Tw_{\phi} & = -3(6 + \cos\phi)/(2(7 + 3 \cos\phi)) \\
   \sigma & = -3(15 - 2 \cos \phi)/(53 - 12\cos\phi) \end{array}
   \right\} \mathrm{for} \; \psi_2 \; \mathrm{with} \; m_1 = 1, m_2 = 0, c = 2/3.
   \label{ex2twist}
\end{equation}
The screw number of this wave is $-1,$ since the phase structure of the summand with larger coefficient dominates.
Only $Tw_{\rm{ell}}$ integrates to this integer; the others give irrational numbers, and this is the case for other choices of $m_1, m_2, c.$
The integrability of $Tw_{\rm{ell}}$ may be explained by the fact that it is the derivative of the difference of angles (\ref{eq:twell}), the others are averages of an angle, and taking the average does not commute with integrating around the dislocation loop. 
Thus, of all the various twists considered, only $Tw_{\rm{ell}}$ can be used as a topological twist.

The twirl is
\begin{equation}
\fl   tw_{\phi} = -(8 - 6\cos\phi)/(25 - 24 \cos\phi)
   \qquad \mathrm{for} \; \psi_2\; \mathrm{with} \; m_1 = 1, m_2 = 0, c = 2/3,
   \label{ex2twirl}
\end{equation}
and, by (\ref{ex2twist}) and (\ref{eq:twell}), $tw_{\chi} = 1+tw_{\phi}.$ 
The topological twirl around a closed loop is also quantised, equal to the number of rotations of the anisotropy ellipse around the loop, although it only needs to undergo a half turn to return to itself smoothly \cite{nye:natural,dennis:polarization}. 
Integrating $tw_{\phi}$ around the loop gives a topological twirl of $-1/2:$ with respect to the anisotropy ellipse, the loop is a M{\"o}bius band. 
The sense of topological twirl and topological twist are independent.

A natural question to ask is whether the topology of twirl gives an analogue of the twisted loop theorem.
In fact, it does: the ellipse rotation around a loop is related to the {\itshape anisotropy C line index} enclosed by the loop. 
Anisotropy C lines, where the anisotropy ellipse is circular, occur when $\nabla \psi \cdot \nabla \psi$ is zero; they correspond to phase singularities of this complex scalar, whose phase is twice the rectifying phase $ \chi_{\rm{r}}.$ 
The anisotropy C line index therefore is half this phase singularity strength, and is equal to the integral of $tw_{\chi}$ around the loop, divided by $2\pi.$ 
This fact, together with the twisted loop theorem and the definition (\ref{eq:twell}) of $Tw_{\rm{ell}},$ leads to the {\itshape twirling loop theorem}, which may be stated as follows.

The anisotropy C line index threading a closed dislocation loop (in units of $1/2$) is equal to the topological twirl minus the topological twist of the loop, i.e. the number of rotations of the anisotropy ellipse around the dislocation, minus the number of rotations of the phase structure, in a right handed sense with respect to the dislocation strength.

This gives a topological role to anisotropy C lines. 
Anisotropy L lines, where $\omega = 0$ (that is, where all the phase surfaces share a common normal), govern the reconnection of dislocation lines \cite{bd:333, nye:airy, nye:local}.

\section{Twist and twirl in isotropic random waves}\label{sec:random}

As a final example, twist and twirl are considered in 3-dimensional isotropic random waves, that is, superpositions of plane waves with isotropic random directions and phases, whose ellipse anisotropy statistics have been calculated \cite{bd:321, dennis:thesis}. 
These random waves might occur in monochromatic waves in a large, chaotic cavity, or (a scalar caricature of) black-body radiation; the wave dislocations form a complicated tangle.
Nevertheless, this random wave model has many statistical symmetries, the averages being spatially and temporally invariant.

The details of the calculations are omitted, but the calculations are reasonably straightforward, following the methods of \cite{bd:321,dennis:thesis}, in which statistics of many geometrical properties of dislocations were calculated. 
The calculations proceed by taking advantage of the fact that $\psi$ and its derivatives have gaussian distributions, the details depending on the power spectrum of the waves considered. 
For all of the probability density functions, the twists and twirls are in units of the {\itshape characteristic twist} $Tw_{\rm{c}} = (k_4/5k_2)^{1/2},$ where $k_n$ is the $n$th moment of the power spectrum; for monochromatic waves of wavelength $\lambda,$ this is $2\pi/\sqrt{5} \lambda.$

\begin{figure}
\begin{center}
\includegraphics*[width=8cm]{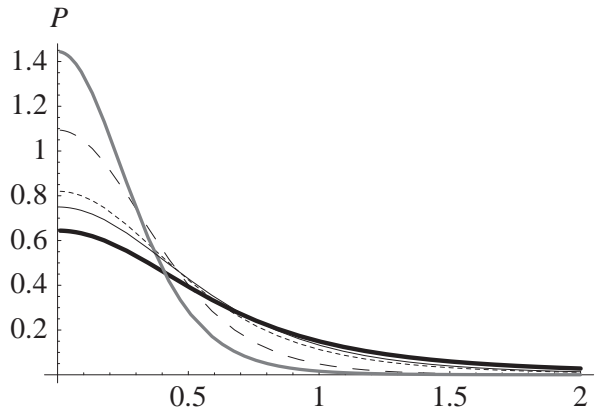}
\end{center}
    \caption{The probability density functions of the different twist and twirl measures. In ascending order of $y$-axis interception, they are: twirl $tw$ (thick black line), helicoid twist $Tw(\chi_0)$ (thin black line), azimuth-averaged twist $Tw_{\phi}$ (dotted line), screwness $\sigma$ (dashed line), and ellipse-averaged twist $Tw_{\rm{ell}}$ (thick grey line). The distributions are all in units of characteristic twist $Tw_{\rm{c}}.$}
    \label{fig:dist}
\end{figure}

All measures of twist and twirl discussed here ($Tw(\chi_0)$ in (\ref{eq:twch}), $Tw_{\phi}$ in (\ref{eq:twph}), $\sigma$ in (\ref{eq:screwness}), $tw_{\phi}$ in (\ref{eq:twphi}), $tw_{\chi}$ in (\ref{eq:twchi}), and $Tw_{\mathrm{ell}}$ in (\ref{eq:twell})) involve some combination of first derivatives of the dislocation with their dislocation directional derivative, divided by another function involving first derivatives only.
In the calculation, the gaussian-distributed second derivatives are integrated first, giving a function involving $\omega$ and $G \equiv |\nabla \psi|^2,$ which is then integrated using the probability distributions described in \cite{bd:321, dennis:thesis}.

The probability distribution functions of the various twists and twirls are found to be 
\begin{eqnarray}
    P_{Tw(\chi_0)}(t) & = & \frac{3}{4} \frac{1}{(1 + t^2)^{5/2}}, \nonumber \\
    P_{\sigma}(t) & = & \frac{35}{32(1+t^2)^{9/2}},
    \nonumber \\
    P_{Tw_{\phi}}(t) & = & \frac{3}{32 t^4}\left(2-
    \frac{(2+7t^2)}{(1+t^2)^{7/2}}\right),
    \nonumber \\   
    P_{Tw_{\rm{ell}}}(t) & = & \frac{1}{8} \left(\frac{\sqrt{2}(11+ 8 t^2 +
    32 t^4)}{(1+2t^2)^{7/2}} -
    \frac{4}{(1+t^2)^{3/2}}\right),
    \nonumber \\
    P_{tw}(t) & = & \frac{1}{32t^4 (1+t^2)^{5/2}} \left( (2+t^2)(3 + 3t^2 +
    8t^4) E \left(\frac{t^2}{1+t^2}\right)\right.
    \nonumber \\
    & &  \qquad \qquad \qquad \left.-2(3+3t^2 + 2t^4)
    K\left(\frac{t^2}{1+t^2}\right) \right),
    \label{eq:twpdfs}
\end{eqnarray}
where $E,K$ represent the complete elliptic integrals of first and second kinds \cite{as:handbook}.
These are plotted in \ref{fig:dist}; they are all symmetric, since right-handed and left-handed screw dislocations are equally weighted in the ensemble.
All of the twist distributions have power law tails.
The fluctuations of the $P_{tw}$ are the largest (the second moment does not converge). 
The fluctuations of $Tw_{\rm{ell}}$ are the smallest (second moment is $\log \sqrt{2} - 1/4$), providing further support for the preference of $Tw_{\rm{ell}}.$

\section{Discussion}

Twist and twirl for anisotropic 3-dimensional wave dislocations are not important for the simplest optical vortices because they are isotropic and therefore not subject to the subtleties described here. 
However, it is important to note that the screw-edge distinction only applies in special cases such as (\ref{eq:edgescrew}), and in more chaotic fields, only local phase geometry may be appealed to.
Various measures of twist and twirl were introduced in \ref{sec:geometry}, their topology was examined in section \ref{sec:closed}, and statistics in \ref{sec:random}. The conclusion from each of these deliberations was that the ellipse-defined twist $Tw_{\rm{ell}}$ (\ref{eq:twell}) has the most desirable properties of the various measures.

Experimental verification of the twirling loop dislocation (\ref{eq:twex}) should be straightforward; with measurements of the phase, the twist and twirl structures of more complicated, curved dislocations may be found, although the full experimental analysis of the twist and twirl of an arbitrary dislocation line is likely to be rather difficult.

It is natural to ask whether there is any curvature structure (i.e. second spatial derivative of $\psi$) transverse to a dislocation, providing a transverse analogue to twist.
This would provide the geometric counterpart to the transverse component of the Burgers vector, as twist is related to the longitudinal part.
For the standard edge dislocation (\ref{eq:edgescrew}), the Burgers vector is normal to the phase surface (mod $\pi$) whose local transverse curvature is zero (a phase saddle in the transverse plane also occurs on this surface); it is tempting to guess this may be the desired structure. 
After some analysis, however, it can be shown that there is no unique phase contour whose local transverse curvature vanishes: generically, there are either one or three phase contours with this property.
The number of phase contours in question is given by the number of real roots of a certain cubic in $\tan \phi,$ reminiscent of the number of straight lined terminating on a line field singularity (lemon versus star and monstar) \cite{bh:60,dennis:polarization}.
An example of a solution to the wave equation with three phase lines of vanishing transverse curvature is 
\begin{equation}
   \psi = \left( x + \rmi y + \rmi k (x^2-y^2/2)/2\right) \exp(\rmi k z).
   \label{eq:cubic}
\end{equation}
Phase patterns for dislocations in two dimensions with this property are plotted in \cite{nhh:tides}, figure 6 (f).
It is not certain, therefore, whether local dislocation geometry can give a transverse direction to a Burgers vector.

The philosophy of studying phase singularities is that they give information about the global phase structure: they are a 3-dimensional `skeleton' for the entire field. 
Twist is an important, intrinsic property of phase singularities, and shows how this information may be gleaned from the local dislocation morphology; closed screw dislocation loops indicate nontrivial wavefront topology, as in figure \ref{fig:noncompact}. 
Phase anisotropy gives rise to further structures, such as twirl, and anisotropy C and L lines.
A possibility of a further extension is analysing the gradient of the anisotropy scalar $\nabla \psi \cdot \nabla \psi$ itself: its phase singularities (the anisotropy C lines) themselves can be twisted, twirling, and therefore related to even higher singular morphologies. 
The twirling loop theorem shows how the first members of this hierarchy of anisotropy structures are coupled.

\section*{Acknowledgements}
I am grateful to Michael Berry, John Hannay and John Nye for many interesting and stimulating discussions. This work was supported by the Leverhulme Trust.

\section*{References}


\end{document}